\documentclass[twocolumn,showpacs,preprintnumbers,amsmath,amssymb]{revtex4}

\usepackage{graphicx}
\usepackage{dcolumn}
\usepackage{bm}

\begin{document}

\title{\large Two path transport measurements on a triple quantum dot}

\author{M. C. Rogge}
\email{rogge@nano.uni-hannover.de}
\author{R.~J. Haug}
\affiliation{Institut f\"ur Festk\"orperphysik, Leibniz
Universit\"at Hannover, Appelstr. 2, 30167 Hannover, Germany}

\date{\today}

\begin{abstract}
We present an advanced lateral triple quantum dot made by local
anodic oxidation. Three dots are coupled in a starlike geometry
with one lead attached to each dot thus allowing for multiple path
transport measurements with two dots per path. In addition charge
detection is implemented using a quantum point contact. Both in
charge measurements as well as in transport we observe clear
signatures of states from each dot. Resonances of two dots can be
established allowing for serial transport via the corresponding
path. Quadruple points with all three dots in resonance are
prepared for different electron numbers and analyzed concerning
the interplay of the simultaneously measured transport along both
paths.
\end{abstract}

\maketitle

Since observation and manipulation on the submicron scale were
made possible some decades ago, a huge variety of lateral
nanostructures on semiconductors has been developed and
investigated to gain access to quantummechanical quantities. By
now even zerodimensional quantum dots \cite{Kouwenhoven-97}, so
called artificial atoms, have been investigated intensively as
they were proposed as crucial elements for quantum computing
\cite{Loss-98}. Next to transport measurements the combination
with quantum point contacts (QPC) has allowed for charge detection
gaining access to new quantities like electron number or coupling
symmetries e.g. \cite{Field-93,Nemutudi-04,Schleser-05,Rogge-05}.
For several years now double quantum dots with two dots combined
to artificial molecules are subject to extensive research
\cite{Wiel-03}. Coupling phenomena and the roll of electronic
spins have been studied in parallel configurations with each dot
connected to both leads \cite{Holleitner-01,Rogge-03} as well as
in serial systems e.g.
\cite{Pioro-Ladriere-03,Elzerman-03,Petta-04,Huttel-05}.

Despite these successes and although there are interesting
theoretical predictions
\cite{Saraga-03,Michaelis-06,Groth-06,Emary-07}, lateral triple
quantum dots have almost not been investigated so far. Some
experiments were done in the early nineties
\cite{Haug-95,Waugh-95}. Three experiments have been published
recently with geometries based on electron beam lithography. Vidan
et al. \cite{Vidan-04} investigated a serial double quantum dot
with a side coupled third dot as a quantum box. Gaudreau et al.
\cite{Gaudreau-06} observed charge rearrangements on a ringlike
triple quantum dot in a system originally designed for double
quantum dots. Schr\"oer et al. \cite{Schroer-07} created a system
with three dots in a row.

In this letter we present a different geometry for a lateral
triple quantum dot. We created a starlike system with each dot
placed next to the other two. In contrast to the formerly
published works we have three leads connected to our system, one
for each quantum dot. Thus we can simultaneously measure transport
via different paths with only two dots per path.

\begin{figure}
 \includegraphics{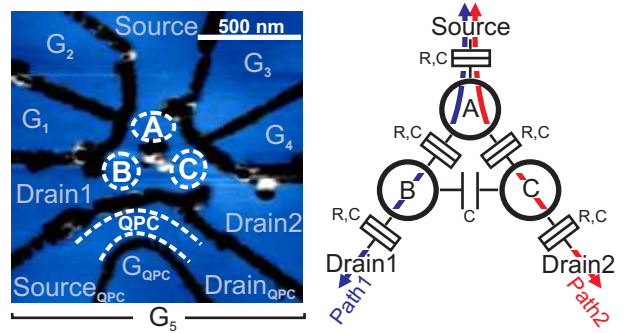}
 \caption{left: Colorized AFM image of our triple dot device defined by oxide lines
 (dark). Three quantum dots A, B, C are placed in the center of the
 device such that each dot is connected via tunnelling barriers to
 the other two. Each dot has its own lead that can serve as a
 source or drain contact. The displayed setup with Source at dot
 A, Drain1 at B and Drain2 at C is used for two path transport
 measurements. Four gates G$_\mathrm{1}$ to G$_\mathrm{4}$ control the potentials of the
 dots and the barriers. A quantum point contact (QPC) is placed next
 to dots B and C for charge detection. It has its own source and
 drain leads (Source$_\mathrm{QPC}$, Drain$_\mathrm{QPC}$) and a tuning gate G$_\mathrm{QPC}$. The
 complete QPC can be used as an additional gate (G$_\mathrm{5}$) with the QPC
 still working. right: Schematic for the triple dot setup. The three dots are
 coupled via tunneling barriers (R, C) to the leads and to each other,
 dots B and C are coupled capacitively only (C). Transport is measured
 via two paths, along dots A, B (path 1) and A, C (path 2).}
 \label{fig1}
\end{figure}

To enable charge detection we extend our device by a QPC making
the setup more flexible. With this variety of features we decided
to use an atomic force microscope (AFM) to built this unique setup
as this technique provides the same functionality with less gates
involved compared to devices made by split gate technique with
ebeam lithography. Therefore as far as we know this is the only
lateral triple quantum dot made with local anodic oxidation (LAO)
\cite{Ishii-95,Keyser-00}.

Using LAO on a GaAs/AlGaAs heterostructure oxide lines are created
shown in black in the AFM-image of the device in Fig.\ref{fig1},
left. Three dots A, B, C (see dashed circles) are defined in a
starlike setup with tunnelling barriers in between. Each dot is
connected to a "personal" lead used as the source or drain
contacts. Four gates G$_\mathrm{1}$ to G$_\mathrm{4}$ are used to
tune the coupling to the leads and the interdot coupling. Due to
the small dimensions the gates do not work independently and thus
a fine balance of all four gate voltages is necessary to operate
the system. For charge detection a QPC (dashed lines) is placed
below dots B and C with its own source and drain leads
(Source$_\mathrm{QPC}$, Drain$_\mathrm{QPC}$) and an additional
gate G$_\mathrm{QPC}$ to tune the conductance of the QPC. The
complete QPC can be used as another gate G$_\mathrm{5}$ for the
triple dot with the charge detection still working.

The measurements were performed in a He3/He4-dilution refrigerator
at a base temperature of 15~mK. AC and DC voltages were applied to
one lead of the triple dot called Source while the differential
conductance G through the device was measured on the other two
leads called Drain1 and Drain2 individually using two lock-in
setups. The QPC was operated with a DC voltage applied to
Source$_\mathrm{QPC}$ and the DC current measured at
Drain$_\mathrm{QPC}$. In the following the device is connected as
shown in Fig. \ref{fig1}: the source contact is connected to dot
A, Drain1 is placed at dot B and Drain2 at dot C. Thus transport
can be measured in two parallel paths as shown in the schematic on
the right of Fig. \ref{fig1} with dots A and B in series along
path 1 and dots A and C in series along path 2.

\begin{figure}
 \includegraphics{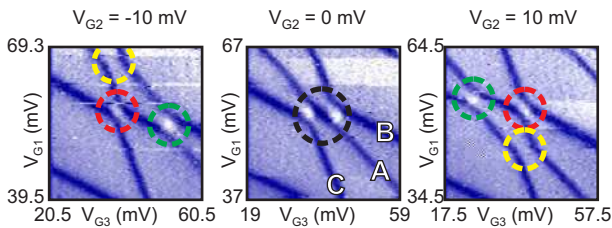}
 \caption{Charge detection of triple dot states. The differentiated
 current through the QPC $dI_{QPC}/dV_{G3}$ is plotted as a function of two gate voltages
 for different values of V$_\mathrm{G2}$ with V$_\mathrm{G4}=0$. In each image dark features are
 visible with different slopes. They denote states from all
 three dots (marked for V$_\mathrm{G2} = 0$~mV). Anticrossings appear for
 resonances of two of the three dots (marked with circles, red for
 resonance of B and C, green for B and A, yellow for C and A).
 With V$_\mathrm{G2}$ those can be shifted to establish a triple dot resonance
 (black circle for V$_\mathrm{G2} = 0$~mV).}
 \label{fig2}
\end{figure}

First the device is studied in the closed regime where no current
is detectable along both paths. Still the charging of the device
is detectable using the QPC. This is shown in Fig. \ref{fig2}. The
derivative of the QPC current is plotted as a function of gates
G$_\mathrm{3}$ and G$_\mathrm{1}$ for three different voltages at
gate G$_\mathrm{2}$. In each measurement dark lines are visible
denoting charging events in the system. As influences from the
gate voltages are different for each dot the charging events show
three different slopes, one for each dot. The lines with the
lowest slope belong to dot B, those with the largest slope belong
to dot C. The lines with intermediate slopes denote charging on
dot A (see Fig. \ref{fig2}, V$_\mathrm{G2}=0$~mV). Where the lines
meet anticrossings are found with two dots in resonance. In Fig.
\ref{fig2} at V$_\mathrm{G2}=-10$~mV those anticrossings are
visible for resonance between dot A and dot B (green circle, the
chemical potentials for dot A and dot B are equal,
$\mu_\mathrm{NA}=\mu_\mathrm{NB}$ with NA and NB the electron
numbers on both dots), dot A and dot C (yellow circle,
$\mu_\mathrm{NA}=\mu_\mathrm{NC}$) and dot B and dot C (red
circle, $\mu_\mathrm{NB}=\mu_\mathrm{NC}$). Although no tunnel
coupling is possible between B and C \cite{bc} there is still a
huge capacitive coupling demonstrating the close vicinity of the
two dots. At these anticrossings bright features are visible due
to charge transitions from one dot to another without changing the
total charge of the system. For V$_\mathrm{G2}=-10$~mV in Fig.
\ref{fig2} all three anticrossings are separated by a few mV. For
V$_\mathrm{G2}=0$~mV these anticrossings coincide (black circle)
thus showing the resonance condition for all three dots
($\mu_\mathrm{NA}=\mu_\mathrm{NB}=\mu_\mathrm{NC}$). With
increasing V$_\mathrm{G2}$ the resonances are shifted further and
the three dot resonance condition is lifted again
(V$_\mathrm{G2}=10$~mV). Thus the existence of three coupled
quantum dots with tunable resonance conditions is demonstrated.
Furthermore, dot C can be emptied to zero electrons as the line
visible for dot C is the last line detected. No further line
corresponding to C appears when decreasing the gate voltages. The
line visible denotes charging with the first electron
($\mu_\mathrm{NC}=\mu_\mathrm{1C}$).

\begin{figure}
 \includegraphics{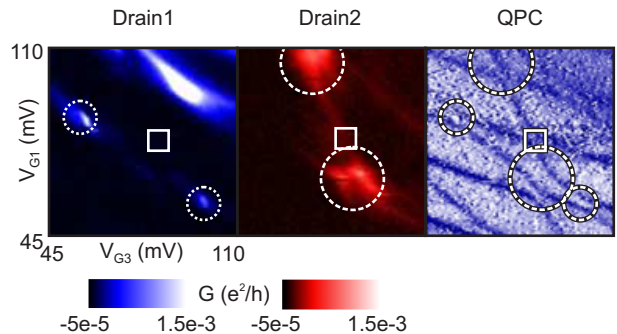}
 \caption{Differential conductance G measured at Drain1 (blue) and Drain2 (red) with the corresponding charge measurement as a
 function of V$_\mathrm{G3}$ and V$_\mathrm{G1}$ with G$_\mathrm{4}$ at -100~mV. Along both paths
 spots of finite G show serial transport due to resonances of the corresponding two
 dots (marked with small circles along path 1 and big circles along path
 2). These spots correspond to anticrossings as shown in the QPC
 measurement. Blue spots denote resonance of dots A and B, red
 spots appear for resonance of dots A and C. The QPC measurement
 shows further resonances of dots B and C (square). Those are not
 visible in transport.}
 \label{fig3}
\end{figure}

Adjusting the gate voltages and opening the barriers finite
transport through the dots becomes measurable. This is shown in
Fig. \ref{fig3}. Charge diagrams are recorded sweeping gates
G$_\mathrm{3}$ and G$_\mathrm{1}$ as in Fig. \ref{fig2}. Next to
the charge detection the differential conductance G is measured
along paths 1 and 2. The measurement along path 1 is plotted in
blue, the one along path 2 in red. Along both paths spot like
features are visible, some of them marked with circles. Comparing
these features with the QPC measurement one finds that they
correspond to anticrossings of quantum dot states. The spots
measured at Drain1 (small circles) correspond to resonances
between dots A and B. The features measured at Drain2 (big
circles) appear due to resonances between dots A and C. The latter
ones are slightly split due to the strong interdot coupling of A
and C. In both paths at least two spots are visible meaning that
these are resonances for different electron numbers. The two
marked blue spots appear for the same state on dot A but for two
consecutive states on dot B. Thus at one spot dot B is charged
with an even number of electrons and at the other resonance with
an odd number. Similar properties account for the red features.
Both spots appear for the same state on dot C, which is charging
with the second electron here (transport for the first electron
can be measured for different gate voltages). Two consecutive
states on dot A are involved one of them with even, the other with
odd electron numbers. The charge signal at the QPC shows a third
group of anticrossings exemplarily marked with a square. Those
correspond to resonances between dots B and C. Comparing these
features with the measurements along both paths it becomes obvious
that there are no corresponding features in transport. This
remarkably shows the functionality of the two path setup with a
common source contact and two drain contacts. In each path the two
dots placed there must be in resonance to allow for serial
transport. Resonances of two dots in different paths are not
sufficient to generate finite differential conductance.

\begin{figure}
 \includegraphics{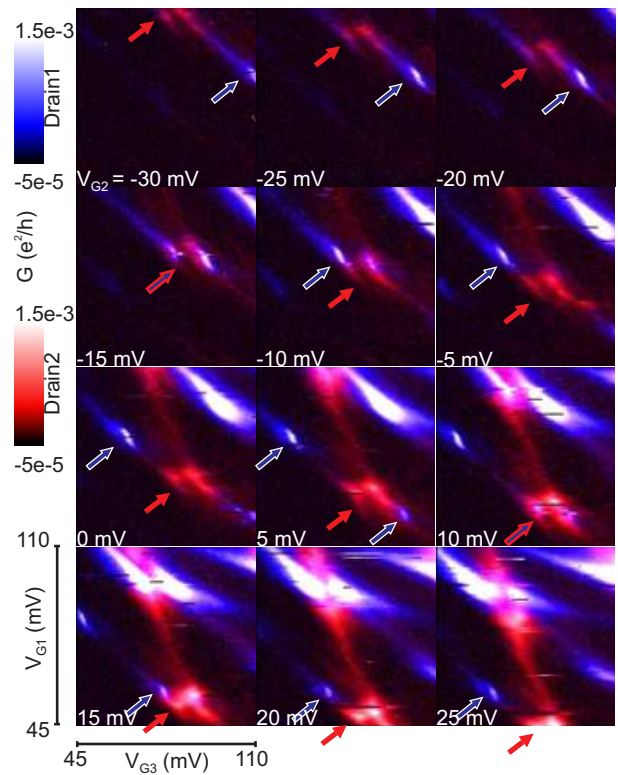}
 \caption{Differential conductance G along
 both paths (along dots A,C in red, along dots A, B in blue) as a
 function of V$_\mathrm{G3}$ and V$_\mathrm{G1}$ while stepping V$_\mathrm{G2}$ from -30~mV (upper left) to 25~mV (lower right). V$_\mathrm{G4}$ is set to -100~mV. Along both paths
 spots of finite G show serial transport due to resonances of the corresponding two dots, some of them marked with accordingly colored arrows.
 With V$_\mathrm{G2}$ those are shifted and triple resonances are established.
 Those appear for V$_\mathrm{G2}=-15$~mV and V$_\mathrm{G2}=10$~mV. Further triple resonances are
 found for higher voltages (e.g. for V$_\mathrm{G2}=20$~mV).}
 \label{fig4}
\end{figure}

With the ability of measuring conductance along two paths
simultaneously but separately our device enables a novel way of
data analysis by plotting the data in advanced color scale plots.
This is demonstrated in 12 consequent plots in Fig. \ref{fig4},
that show both signals simultaneously in one plot but separated by
different colors (red for path 1, blue for path 2). Thus this way
of plotting suits perfectly the original data recording. Features
in both paths can be compared immediately and assigned to the
appropriate path. For example the strong blue to white features in
the upper part of the last three plots are split into two. One can
directly see that this comes from a red feature that splits the
blue features due to interdot coupling.

The main aspect of the 12 consequent plots in Fig. \ref{fig4} is
to study the way triple resonances appear in transport. Similar as
in Fig. \ref{fig2} the resonances visible in transport can be
shifted to establish resonances of all three quantum dots. For
each measurement shown in Fig. \ref{fig4} V$_\mathrm{G2}$ is set
to a fixed voltage starting from -30~mV at the upper left and
increased to 25~mV at the lower right. Both paths show spot like
features due to resonance of two dots (color encoded as in Fig.
\ref{fig3}) some of them marked with blue and red arrows. While
the red spots move downwards with increasing V$_\mathrm{G2}$ the
blue spots move to the left. Thus using gate G$_\mathrm{2}$
resonance conditions of all three dots can be created. The two
spots marked with a red and a blue arrow for
V$_\mathrm{G2}=-30$~mV (chemical potentials
$\mu_\mathrm{NA}=\mu_\mathrm{NB}$ and
$\mu_\mathrm{NA}=\mu_\mathrm{2C}$) for example approach each other
with increasing V$_\mathrm{G2}$. At V$_\mathrm{G2}=-15$~mV both
spots have merged, simultaneous transport via path 1 and path 2 is
detected, the three dots are in resonance
($\mu_\mathrm{NA}=\mu_\mathrm{NB}=\mu_\mathrm{2C}$). Thereby the
blue resonance is split into two resonances due to the strong
anticrossing for the red spot (a similar splitting of the red spot
is almost not observable due to the much weaker anticrossing for
the blue spot). A further increase of V$_\mathrm{G2}$ moves the
blue and red spots apart again. The same happens with the red spot
and another blue spot coming into resonance at
V$_\mathrm{G2}=10$~mV
($\mu_\mathrm{NA}=\mu_\mathrm{NB-1}=\mu_\mathrm{2C}$). The two
triple resonances differ by one electron on dot B. As mentioned
before further resonances are visible at higher V$_\mathrm{G1}$
($\mu_\mathrm{NA+1}=\mu_\mathrm{NB+1}=\mu_\mathrm{2C}$ at
V$_\mathrm{G2}=20$~mV). Here the electron number on dot A has
changed as well. Similar results were gained stepping
V$_\mathrm{G4}$ instead of V$_\mathrm{G2}$.

Thus not only with charge detection, even in transport we can
detect clear resonances of two quantum dots in two different paths
in combined color plots. Shifting these resonances quadruple
points can be formed with all three dots in resonance. These
quadruple points can be prepared for different electron numbers
creating odd or even configurations on each dot and thus on the
whole triple dot as well. Therefore this device is promising to
verify theoretical predictions published recently for two path
triple quantum dots. A two path triple dot can be used as a spin
entangler \cite{Saraga-03} with a spin singlet formed in dot A for
an even number of electrons. As mentioned before we can prepare an
even number of electrons for each dot. The entangled spins are
then separated and transferred to the two drain leads with one
electron per path thus creating spin entangled currents. Refs.
\cite{Michaelis-06,Groth-06,Emary-07} predict the formation of a
trapped state for electrons entering the triple dot via dots B and
C. A coherent superposition of charge in the two dots can be
created with destructive interference at dot A blocking transport.
Depending on the needed direction of the current flow through the
two paths one could use different setups of source and drain
contacts on our device to establish the appropriate conditions for
both experiments.

In conclusion we have investigated resonances of two and three
quantum dots in transport and with charge detection in a lateral
triple dot device made with local anodic oxidation. The three dots
are arranged in a starlike geometry with each of them coupled to
the other two. Three leads, one for each dot, allow for
simultaneous transport measurements via different paths. States
from all three dots were detected in charge measurements showing
anticrossings when two dots come into resonance. Adjusting the
four gate voltages it was possible to establish resonances for all
three dots. This tunability was confirmed in transport
measurements at different gate voltages. Via two paths transport
was measured simultaneously with each path showing resonances of
two dots. Resonance conditions were established with simultaneous
transport via both paths with all three dots in resonance. The
formation of these quadruple points was analyzed for both paths
simultaneously in combined color scale plots for different
electron numbers.

This work has been supported by BMBF via nanoQUIT.


\newpage


\begin{thebibliography}{20}
\expandafter\ifx\csname natexlab\endcsname\relax\def\natexlab#1{#1}\fi
\expandafter\ifx\csname bibnamefont\endcsname\relax
  \def\bibnamefont#1{#1}\fi
\expandafter\ifx\csname bibfnamefont\endcsname\relax
  \def\bibfnamefont#1{#1}\fi
\expandafter\ifx\csname citenamefont\endcsname\relax
  \def\citenamefont#1{#1}\fi
\expandafter\ifx\csname url\endcsname\relax
  \def\url#1{\texttt{#1}}\fi
\expandafter\ifx\csname urlprefix\endcsname\relax\def\urlprefix{URL }\fi
\providecommand{\bibinfo}[2]{#2}
\providecommand{\eprint}[2][]{\url{#2}}



\bibitem[{\citenamefont{Kouwenhoven et~al.}(1997)\citenamefont{Kouwenhoven,
  Marcus, McEuen, Tarucha, Westervelt, and Wingreen}}]{Kouwenhoven-97}
\bibinfo{author}{\bibfnamefont{L.~P.} \bibnamefont{Kouwenhoven}},
  \bibinfo{author}{\bibfnamefont{C.~M.} \bibnamefont{Marcus}},
  \bibinfo{author}{\bibfnamefont{P.~L.} \bibnamefont{McEuen}},
  \bibinfo{author}{\bibfnamefont{S.}~\bibnamefont{Tarucha}},
  \bibinfo{author}{\bibfnamefont{R.~M.} \bibnamefont{Westervelt}},
  \bibnamefont{and} \bibinfo{author}{\bibfnamefont{N.~S.}
  \bibnamefont{Wingreen}}, in \emph{\bibinfo{booktitle}{Mesoscopic Electron
  Transport}}, edited by \bibinfo{editor}{\bibfnamefont{L.~L.}
  \bibnamefont{Sohn}}, \bibinfo{editor}{\bibfnamefont{L.~P.}
  \bibnamefont{Kouwenhoven}}, \bibnamefont{and}
  \bibinfo{editor}{\bibfnamefont{G.}~\bibnamefont{Sch\"o{}n}}
  (\bibinfo{publisher}{Kluwer}, \bibinfo{address}{Dordrecht},
  \bibinfo{year}{1997}), vol. \bibinfo{volume}{345} of
  \emph{\bibinfo{series}{Series E}}, pp. \bibinfo{pages}{105--214}.

\bibitem[{\citenamefont{Loss and DiVincenzo}(1998)}]{Loss-98}
\bibinfo{author}{\bibfnamefont{D.}~\bibnamefont{Loss}} \bibnamefont{and}
  \bibinfo{author}{\bibfnamefont{D.~P.} \bibnamefont{DiVincenzo}},
  \bibinfo{journal}{Phys. Rev. A} \textbf{\bibinfo{volume}{57}},
  \bibinfo{pages}{120} (\bibinfo{year}{1998}).

  \bibitem[{\citenamefont{Field et~al.}(1993)\citenamefont{Field, Smith, Pepper, Richie, Frost, Jones, and Hasko}}]{Field-93}
\bibinfo{author}{\bibfnamefont{M.} \bibnamefont{Field}},
\bibinfo{author}{\bibfnamefont{C.~G.}~\bibnamefont{Smith}},
\bibinfo{author}{\bibfnamefont{M.}~\bibnamefont{Pepper}},
\bibinfo{author}{\bibfnamefont{D.~A.}~\bibnamefont{Ritchie}},
\bibinfo{author}{\bibfnamefont{J.~E.~F.}~\bibnamefont{Frost}},
\bibinfo{author}{\bibfnamefont{G.~A.~C.}~\bibnamefont{Jones}},
\bibnamefont{and} \bibinfo{author}{\bibfnamefont{D.~G.} \bibnamefont{Hasko}},
  \bibinfo{journal}{Phys. Rev. Lett.} \textbf{\bibinfo{volume}{70}},
  \bibinfo{pages}{1311} (\bibinfo{year}{1993}).

  \bibitem[{\citenamefont{Nemutudi et~al.}(2004)\citenamefont{Nemutudi, Kataoka, Ford, Appleyard, Pepper, Ritchie, and Jones}}]{Nemutudi-04}
\bibinfo{author}{\bibfnamefont{R.} \bibnamefont{Nemutudi}},
\bibinfo{author}{\bibfnamefont{M.}~\bibnamefont{Kataoka}},
\bibinfo{author}{\bibfnamefont{C.~J.~B.}~\bibnamefont{Ford}},
\bibinfo{author}{\bibfnamefont{N.~J.}~\bibnamefont{Appleyard}},
\bibinfo{author}{\bibfnamefont{M.}~\bibnamefont{Pepper}},
\bibinfo{author}{\bibfnamefont{D.~A.}~\bibnamefont{Ritchie}},
\bibnamefont{and} \bibinfo{author}{\bibfnamefont{G.~A.~C.} \bibnamefont{Jones}},
  \bibinfo{journal}{J. Appl. Phys.} \textbf{\bibinfo{volume}{95}},
  \bibinfo{pages}{2557} (\bibinfo{year}{2004}).

  \bibitem[{\citenamefont{Schleser et~al.}(2005)\citenamefont{Schleser, Ruh, Ihn, Ensslin, Driscoll, and Gossard}}]{Schleser-05}
\bibinfo{author}{\bibfnamefont{R.} \bibnamefont{Schleser}},
\bibinfo{author}{\bibfnamefont{E.}~\bibnamefont{Ruh}},
\bibinfo{author}{\bibfnamefont{T.}~\bibnamefont{Ihn}},
\bibinfo{author}{\bibfnamefont{K.}~\bibnamefont{Ensslin}},
\bibinfo{author}{\bibfnamefont{D.~C.}~\bibnamefont{Driscoll}},
\bibnamefont{and} \bibinfo{author}{\bibfnamefont{A.~C.} \bibnamefont{Gossard}},
  \bibinfo{journal}{Phys. Rev. B} \textbf{\bibinfo{volume}{72}},
  \bibinfo{pages}{035312} (\bibinfo{year}{2005}).

\bibitem[{\citenamefont{Rogge et~al.}(2005)\citenamefont{Rogge, Harke, Fricke, Hohls, Reinwald, Wegscheider, and Haug}}]{Rogge-05}
\bibinfo{author}{\bibfnamefont{M.~C.} \bibnamefont{Rogge}},
\bibinfo{author}{\bibfnamefont{B.}~\bibnamefont{Harke}},
\bibinfo{author}{\bibfnamefont{C.}~\bibnamefont{Fricke}},
\bibinfo{author}{\bibfnamefont{F.} \bibnamefont{Hohls}},
\bibinfo{author}{\bibfnamefont{M.}~\bibnamefont{Reinwald}},
  \bibinfo{author}{\bibfnamefont{W.}~\bibnamefont{Wegscheider}},
  \bibnamefont{and} \bibinfo{author}{\bibfnamefont{R.~J.}~\bibnamefont{Haug}},
  \bibinfo{journal}{Phys. Rev. B} \textbf{\bibinfo{volume}{72}},
  \bibinfo{pages}{233402} (\bibinfo{year}{2005}).

\bibitem[{\citenamefont{van~der Wiel et~al.}(2003)\citenamefont{van~der Wiel,
  Franceschi, Elzerman, Fujisawa, Tarucha, and Kouwenhoven}}]{Wiel-03}
\bibinfo{author}{\bibfnamefont{W.~G.} \bibnamefont{van~der Wiel}},
  \bibinfo{author}{\bibfnamefont{S.~D.} \bibnamefont{Franceschi}},
  \bibinfo{author}{\bibfnamefont{J.~M.} \bibnamefont{Elzerman}},
  \bibinfo{author}{\bibfnamefont{T.}~\bibnamefont{Fujisawa}},
  \bibinfo{author}{\bibfnamefont{S.}~\bibnamefont{Tarucha}}, \bibnamefont{and}
  \bibinfo{author}{\bibfnamefont{L.~P.} \bibnamefont{Kouwenhoven}},
  \bibinfo{journal}{Rev. Mod. Phys.} \textbf{\bibinfo{volume}{75}},
  \bibinfo{pages}{1} (\bibinfo{year}{2003}).


\bibitem[{\citenamefont{Holleitner et~al.}(2001)\citenamefont{Holleitner,
  Decker, Qin, Eberl, and Blick}}]{Holleitner-01}
\bibinfo{author}{\bibfnamefont{A.~W.} \bibnamefont{Holleitner}},
  \bibinfo{author}{\bibfnamefont{C.~R.} \bibnamefont{Decker}},
  \bibinfo{author}{\bibfnamefont{H.}~\bibnamefont{Qin}},
  \bibinfo{author}{\bibfnamefont{K.}~\bibnamefont{Eberl}}, \bibnamefont{and}
  \bibinfo{author}{\bibfnamefont{R.~H.} \bibnamefont{Blick}},
  \bibinfo{journal}{Phys. Rev. Lett.} \textbf{\bibinfo{volume}{87}},
  \bibinfo{pages}{256802} (\bibinfo{year}{2001}).

\bibitem[{\citenamefont{Rogge et~al.}(2003)\citenamefont{Rogge, F\"uhner, Keyser, Haug, Bichler, Abstreiter, and Wegscheider}}]{Rogge-03}
\bibinfo{author}{\bibfnamefont{M.~C.} \bibnamefont{Rogge}},
\bibinfo{author}{\bibfnamefont{C.}~\bibnamefont{F\"uhner}},
\bibinfo{author}{\bibfnamefont{U.~F.}~\bibnamefont{Keyser}},
\bibinfo{author}{\bibfnamefont{R.~J.} \bibnamefont{Haug}},
\bibinfo{author}{\bibfnamefont{M.}~\bibnamefont{Bichler}},
  \bibinfo{author}{\bibfnamefont{G.}~\bibnamefont{Abstreiter}},
  \bibnamefont{and} \bibinfo{author}{\bibfnamefont{W.}~\bibnamefont{Wegscheider}},
  \bibinfo{journal}{Appl. Phys. Lett.} \textbf{\bibinfo{volume}{83}},
  \bibinfo{pages}{1163} (\bibinfo{year}{2003}).

    \bibitem[{\citenamefont{Pioro-Ladriere et~al.}(2003)\citenamefont{Pioro-Ladriere, Ciorga, Lapointe, Zawadzki, Korkusinski, Hawrylak, and Sachrajda}}]{Pioro-Ladriere-03}
\bibinfo{author}{\bibfnamefont{M.}~\bibnamefont{Pioro-Ladriere}},
\bibinfo{author}{\bibfnamefont{M.} \bibnamefont{Ciorga}},
\bibinfo{author}{\bibfnamefont{J.}~\bibnamefont{Lapointe}},
\bibinfo{author}{\bibfnamefont{P.}~\bibnamefont{Zawadzki}},
\bibinfo{author}{\bibfnamefont{M.}~\bibnamefont{Korkusinski}},
\bibinfo{author}{\bibfnamefont{P.}~\bibnamefont{Hawrylak}},
\bibnamefont{and} \bibinfo{author}{\bibfnamefont{A.~S.}~\bibnamefont{Sachrajda}},
  \bibinfo{journal}{Phys. Rev. Lett.} \textbf{\bibinfo{volume}{91}},
  \bibinfo{pages}{026803} (\bibinfo{year}{2003}).

\bibitem[{\citenamefont{Elzerman et~al.}(2003)\citenamefont{Elzerman, Hanson, Greidanus, Willems van Beveren, De Franceschi, Vandersypen, Tarucha, and Kouwenhoven}}]{Elzerman-03}
\bibinfo{author}{\bibfnamefont{J.~M.} \bibnamefont{Elzerman}},
\bibinfo{author}{\bibfnamefont{R.}~\bibnamefont{Hanson}},
\bibinfo{author}{\bibfnamefont{J.~S.}~\bibnamefont{Greidanus}},
\bibinfo{author}{\bibfnamefont{L.~H.}~\bibnamefont{Willems van Beveren}},
\bibinfo{author}{\bibfnamefont{S.}~\bibnamefont{De Franceschi}},
\bibinfo{author}{\bibfnamefont{L.~M.~K.}~\bibnamefont{Vandersypen}},
\bibinfo{author}{\bibfnamefont{S.}~\bibnamefont{Tarucha}},
\bibnamefont{and} \bibinfo{author}{\bibfnamefont{L.~P.} \bibnamefont{Kouwenhoven}},
  \bibinfo{journal}{Phys. Rev. B} \textbf{\bibinfo{volume}{67}},
  \bibinfo{pages}{161308(R)} (\bibinfo{year}{2003}).

\bibitem[{\citenamefont{Petta et~al.}(2004)\citenamefont{Petta, Johnson, Marcus, Hanson, and Gossard}}]{Petta-04}
\bibinfo{author}{\bibfnamefont{J.~R.} \bibnamefont{Petta}},
\bibinfo{author}{\bibfnamefont{A.~C.}~\bibnamefont{Johnson}},
\bibinfo{author}{\bibfnamefont{C.~M.}~\bibnamefont{Marcus}},
\bibinfo{author}{\bibfnamefont{M.~P.}~\bibnamefont{Hanson}},
\bibnamefont{and} \bibinfo{author}{\bibfnamefont{A.~C.} \bibnamefont{Gossard}},
  \bibinfo{journal}{Phys. Rev. Lett.} \textbf{\bibinfo{volume}{93}},
  \bibinfo{pages}{186802} (\bibinfo{year}{2004}).

\bibitem[{\citenamefont{H\"uttel et~al.}(2005)\citenamefont{Huttel, Ludwig, Lorenz, Eberl, and Kotthaus}}]{Huttel-05}
\bibinfo{author}{\bibfnamefont{A.~K.} \bibnamefont{H\"uttel}},
\bibinfo{author}{\bibfnamefont{S.}~\bibnamefont{Ludwig}},
\bibinfo{author}{\bibfnamefont{H.}~\bibnamefont{Lorenz}},
\bibinfo{author}{\bibfnamefont{K.}~\bibnamefont{Eberl}},
\bibnamefont{and} \bibinfo{author}{\bibfnamefont{J.~P.} \bibnamefont{Kotthaus}},
  \bibinfo{journal}{Phys. Rev. B} \textbf{\bibinfo{volume}{72}},
  \bibinfo{pages}{081310(R)} (\bibinfo{year}{2005}).


\bibitem[{\citenamefont{Saraga et~al.}(2003)\citenamefont{Saraga and Loss}}]{Saraga-03}
\bibinfo{author}{\bibfnamefont{D.~S.} \bibnamefont{Saraga}},
\bibnamefont{and} \bibinfo{author}{\bibfnamefont{D.} \bibnamefont{Loss}},
  \bibinfo{journal}{Phys. Rev. Lett.} \textbf{\bibinfo{volume}{90}},
  \bibinfo{pages}{166803} (\bibinfo{year}{2003}).

\bibitem[{\citenamefont{Michaelis et~al.}(2006)\citenamefont{Michaelis, Emary, and Beenakker}}]{Michaelis-06}
\bibinfo{author}{\bibfnamefont{B.} \bibnamefont{Michaelis}},
\bibinfo{author}{\bibfnamefont{C.}~\bibnamefont{Emary}},
\bibnamefont{and} \bibinfo{author}{\bibfnamefont{C.~W.~J.} \bibnamefont{Beenakker}},
  \bibinfo{journal}{Europhys. Lett.} \textbf{\bibinfo{volume}{73}},
  \bibinfo{pages}{677} (\bibinfo{year}{2006}).

  \bibitem[{\citenamefont{Groth et~al.}(2006)\citenamefont{Groth, Michaelis, and Beenakker}}]{Groth-06}
\bibinfo{author}{\bibfnamefont{C.~W.} \bibnamefont{Groth}},
\bibinfo{author}{\bibfnamefont{B.} \bibnamefont{Michaelis}},
\bibnamefont{and} \bibinfo{author}{\bibfnamefont{C.~W.~J.} \bibnamefont{Beenakker}},
  \bibinfo{journal}{Phys. Rev. B} \textbf{\bibinfo{volume}{74}},
  \bibinfo{pages}{125315} (\bibinfo{year}{2006}).

\bibitem[{\citenamefont{Emary}(2006)\citenamefont{Emary}}]{Emary-07}
\bibinfo{author}{\bibfnamefont{C.}~\bibnamefont{Emary}},
  \bibinfo{journal}{cond-mat/07052934}.

\bibitem[{\citenamefont{Haug}(2004)\citenamefont{Haug}}]{Haug-95}
\bibinfo{author}{\bibfnamefont{R.~J.} \bibnamefont{Haug}},
  \bibinfo{journal}{Electrochimica Acta} \textbf{\bibinfo{volume}{40}},
  \bibinfo{pages}{1283} (\bibinfo{year}{1995}).

\bibitem[{\citenamefont{Waugh et~al.}(1995)\citenamefont{Waugh, Berry, Mar, Westervelt, Campman, and Gossard}}]{Waugh-95}
\bibinfo{author}{\bibfnamefont{F.~R.} \bibnamefont{Waugh}},
\bibinfo{author}{\bibfnamefont{M.~J.}~\bibnamefont{Berry}},
\bibinfo{author}{\bibfnamefont{D.~J.}~\bibnamefont{Mar}},
\bibinfo{author}{\bibfnamefont{R.~M.}~\bibnamefont{Westervelt}},
\bibinfo{author}{\bibfnamefont{K.~L.}~\bibnamefont{Campman}},
\bibnamefont{and} \bibinfo{author}{\bibfnamefont{A.~C.} \bibnamefont{Gossard}},
  \bibinfo{journal}{Phys. Rev. Lett.} \textbf{\bibinfo{volume}{75}},
  \bibinfo{pages}{705} (\bibinfo{year}{1995}).

\bibitem[{\citenamefont{Vidan et~al.}(2004)\citenamefont{Vidan, Westervelt, Stopa, Hanson, and Gossard}}]{Vidan-04}
\bibinfo{author}{\bibfnamefont{A.} \bibnamefont{Vidan}},
\bibinfo{author}{\bibfnamefont{R.~M.}~\bibnamefont{Westervelt}},
\bibinfo{author}{\bibfnamefont{M.}~\bibnamefont{Stopa}},
\bibinfo{author}{\bibfnamefont{M.}~\bibnamefont{Hanson}},
\bibnamefont{and} \bibinfo{author}{\bibfnamefont{A.~C.} \bibnamefont{Gossard}},
  \bibinfo{journal}{Appl. Phys. Lett.} \textbf{\bibinfo{volume}{85}},
  \bibinfo{pages}{3602} (\bibinfo{year}{2004}).

\bibitem[{\citenamefont{Gaudreau et~al.}(2006)\citenamefont{Gaudreau, Studenikin, Sachrajda, Zawadzki, Kam, Lapointe, Korkusinski, and Hawrylak}}]{Gaudreau-06}
\bibinfo{author}{\bibfnamefont{L.} \bibnamefont{Gaudreau}},
\bibinfo{author}{\bibfnamefont{S.~A.}~\bibnamefont{Studenikin}},
\bibinfo{author}{\bibfnamefont{A.~S.}~\bibnamefont{Sachrajda}},
\bibinfo{author}{\bibfnamefont{P.}~\bibnamefont{Zawadzki}},
\bibinfo{author}{\bibfnamefont{A.}~\bibnamefont{Kam}},
\bibinfo{author}{\bibfnamefont{J.}~\bibnamefont{Lapointe}},
\bibinfo{author}{\bibfnamefont{M.}~\bibnamefont{Korkusinski}},
\bibnamefont{and} \bibinfo{author}{\bibfnamefont{P.} \bibnamefont{Hawrylak}},
  \bibinfo{journal}{Phys. Rev. Lett.} \textbf{\bibinfo{volume}{97}},
  \bibinfo{pages}{036807} (\bibinfo{year}{2006}).

\bibitem[{\citenamefont{Schr\"oer et~al.}(2007)\citenamefont{Schr\"oer, Greentree, Gaudreau, Eberl, Hollenberg, Kotthaus, and Ludwig}}]{Schroer-07}
\bibinfo{author}{\bibfnamefont{D.} \bibnamefont{Schr\"oer}},
\bibinfo{author}{\bibfnamefont{A.~D.}~\bibnamefont{Greentree}},
\bibinfo{author}{\bibfnamefont{L.}~\bibnamefont{Gaudreau}},
\bibinfo{author}{\bibfnamefont{K.}~\bibnamefont{Eberl}},
\bibinfo{author}{\bibfnamefont{L.~C.~L.}~\bibnamefont{Hollenberg}},
\bibinfo{author}{\bibfnamefont{J.~P.}~\bibnamefont{Kotthaus}},
\bibnamefont{and} \bibinfo{author}{\bibfnamefont{S.} \bibnamefont{Ludwig}},
  \bibinfo{journal}{cond-mat/0703450}.

\bibitem{bc}
With Source at dot B and Drain at A and C we found, that the
barrier connecting dots B and C cannot be opened while keeping the
dots working even with 210~mV applied to gate G$_\mathrm{5}$. The
low tunability originates from the fact that there is only one
large gate (G$_\mathrm{5}$) on that side of the device.


\bibitem[{\citenamefont{Ishii and Matsumoto}(1995)}]{Ishii-95}
\bibinfo{author}{\bibfnamefont{M.}~\bibnamefont{Ishii}} \bibnamefont{and}
  \bibinfo{author}{\bibfnamefont{K.}~\bibnamefont{Matsumoto}},
  \bibinfo{journal}{Jpn. J. Appl. Phys.} \textbf{\bibinfo{volume}{34}},
  \bibinfo{pages}{1329} (\bibinfo{year}{1995}).

\bibitem[{\citenamefont{Keyser et~al.}(2000)\citenamefont{Keyser, Schumacher,
  Zeitler, Haug, and Eberl}}]{Keyser-00}
\bibinfo{author}{\bibfnamefont{U.~F.} \bibnamefont{Keyser}},
  \bibinfo{author}{\bibfnamefont{H.~W.} \bibnamefont{Schumacher}},
  \bibinfo{author}{\bibfnamefont{U.}~\bibnamefont{Zeitler}},
  \bibinfo{author}{\bibfnamefont{R.~J.} \bibnamefont{Haug}}, \bibnamefont{and}
  \bibinfo{author}{\bibfnamefont{K.}~\bibnamefont{Eberl}},
  \bibinfo{journal}{Appl. Phys. Lett.} \textbf{\bibinfo{volume}{76}},
  \bibinfo{pages}{457} (\bibinfo{year}{2000}).


\end{thebibliography}



\end{document}